\documentclass{emulateapj}

\newcommand{\kms}{\ensuremath{{\rm km~s}^{-1}}}
\newcommand{\kev}{\ensuremath{\mbox{keV}}}
\newcommand{\ergs}{\ensuremath{\mbox{erg s}^{-1}}}

\newcommand{\balobjl}{FIRST~J101614.3+520916}
\newcommand{\balobj}{J1016+5209}
\newcommand{\xmmn}{{\it XMM-Newton}}
\shorttitle{XMM-Newton Detection of J101614.3+520916}
\shortauthors{Schaefer et al.}

\begin{document}

\title{{\it XMM-Newton} Detection of the Rare 
FR~II BAL Quasar \\
FIRST J101614.3+520916}

\author{Justin J. Schaefer,\altaffilmark{1,2}
Michael S. Brotherton,\altaffilmark{1}
Zhaohui Shang,\altaffilmark{1}
Michael D. Gregg,\altaffilmark{3,4}
\\
Robert H. Becker,\altaffilmark{3,4}
Sally A. Laurent-Muehleisen,\altaffilmark{3}
Mark Lacy,\altaffilmark{5}
and Richard L. White\altaffilmark{6}
}

\altaffiltext{1}{Department of Physics and Astronomy, University of
Wyoming, Laramie, WY 82072}

\altaffiltext{2}{Department of Astronomy, University of Florida, 
Gainesville, FL 32611}

\altaffiltext{3}{Department of Physics, University of California, Davis,
CA 95616}

\altaffiltext{4}{Institute of Geophysics and Planetary Physics,
Lawrence Livermore National Laboratory, L-413, 7000 East Ave,
Livermore, CA 94550}

\altaffiltext{5}{Spitzer Science Center, California Institute of
Technology, MC 220-6, Pasadena, CA 91125}

\altaffiltext{6}{Space Telescope Science Institute, 3700 San Martin
Drive, Baltimore MD 21218}


% Abstract

\begin{abstract}

We have detected FIRST J101614.3+520916 with the \xmmn\ X-ray
Observatory.  FIRST J101614.3+520916, one of the most extreme
radio-loud, broad absorption line (BAL) quasars so far discovered, is
also a Fanaroff-Riley type II (FR II) radio source.  We find that,
compared to its estimated intrinsic X-ray flux, the observed X-rays
are likely suppressed, and that the observed hardness ratio indicates
significant soft X-ray photons.  This is inconsistent with the simplest model,
a normal quasar spectrum absorbed by a large neutral HI column density, 
which would primarily absorb the softer photons.  More complex models, 
involving partial covering, an ionized absorber, ionized mirror reflection, 
or jet contributions need to be invoked to explain this source.  
The suppressed but soft X-ray emission in this radio-loud BAL quasar is 
consistent with the behavior displayed by other BAL quasars, both 
radio-loud and radio-quiet.

\end{abstract}

\section{Introduction}

About 10--20\% of quasars show broad absorption lines (BAL),
especially in their ultraviolet (UV) spectra.  These absorption
features usually extend to velocities as high as $\sim 10^4$\,\kms\ relative
to the emission lines, indicating high-velocity outflows in the
quasars.  These absorbers have been identified with winds blowing
from an obscuring torus or arising from smaller scales associated with
an accretion disk feeding a super massive black hole.  
The popular orientation model suggests that BAL
quasars are normal quasars viewed along the specific line of sight, or
particularly edge-on, skimming the torus or through a wind
\citep[e.g.,][]{Weym91}.  Although in this picture, quasar radio
properties and BALs would seem to be independent, no radio-loud BAL
quasars were found for a long time.  It remained so until deep radio
surveys like the NRAO VLA Sky Survey \citep[NVSS,][]{Cond98} and FIRST
Bright Quasar Survey \citep[FBQS,][]{Beck95,Greg96,Whit00} were
conducted, surveying large areas to mJy levels, and radio-loud BAL
quasars started to be identified \citep{Beck97,Brot98,Beck00,Meno01,Brot02}.
Still, BAL quasar frequency does drop significantly among the most
radio-loud quasars \citep{Beck01}.
\citet{Beck00} studied 27 BAL quasars from the FBQS sample and found
that they show a wide range of radio spectral indices, from flat to
steep, indicating that a range of orientations is present and
therefore strongly challenging the orientation model.

\begin{figure}
\epsscale{0.9}
\plotone{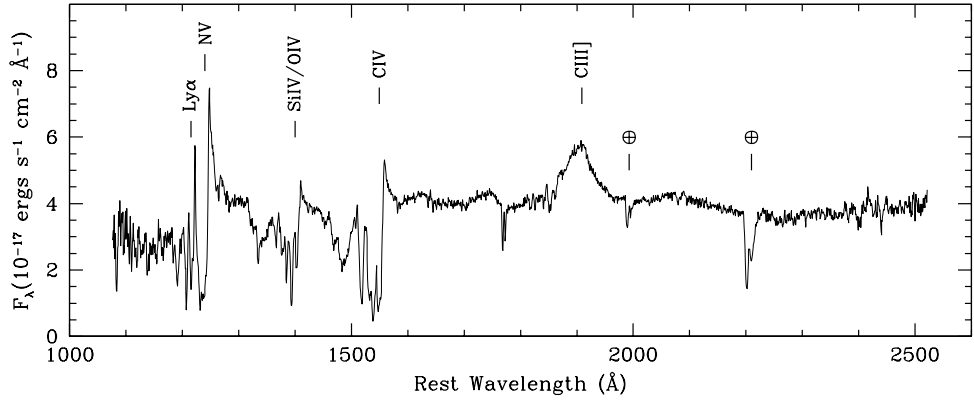}
\caption{Total light spectrum of \balobjl\ from
a spectropolarimetric observation obtained with Keck in January 2000,
showing the broad absorption lines.
Emission-line positions are marked.  Also marked are atmospheric
absorption bands ($\oplus$).
\label{fg:spec}
}
\end{figure}

So far, only a few radio-loud BAL quasars have been studied at X-ray
energies (Brotherton et al. 2005).  \balobjl\ (hereafter \balobj) is
the first confirmed BAL quasar that has also been identified as
radio-loud Fanaroff-Riley type II (FR~II) source \citep{Greg00}.
Figure~\ref{fg:spec} shows the BALs in a rest-frame UV spectrum of
\balobj\ and Table~\ref{tb:para} provides its optical and radio
parameters.  The radio luminosity places it among the extreme end of
radio-loud BAL quasars.  Its double-lobed radio morphology and
luminosity indicate a classic FR~II radio source.  \citet{Greg00}
argue that \balobj\ is a rejuvenated quasar, possibly through a merger
or interaction.  We note that another known radio-loud, FR~II BAL
quasar is LBQS~1138-0126 \citep{Brot02}, and that the double radio lobed
BAL quasar candidate PKS~1004+13 \citep{Will99} has recently been 
confirmed by {\em HST} observation \citep{Will06} as a bone fide 
FR~II BAL quasar.  The other known radio-loud BAL quasars have compact 
structures \citep{Beck00}.

%another radio-loud quasar
%PKS~1004+13 may be a low-luminosity, low-redshift object of similar
%nature \citep{Will99}.  

Observations with the {\it Chandra} X-ray Observatory and \xmmn\ show that
BAL quasars are up to 2 orders of magnitude fainter in X-rays than non-BAL
quasars of the same optical brightness \citep{Gree01, SabHam01,
Gall02,Brot05}.  Available X-ray spectral analyses of radio-quiet BAL quasars 
show that they appear to have normal radio-quiet X-ray photon indices
($\Gamma \approx 2$), partially or totally covered by absorbing
columns of $N_H\leq 10^{23}$cm$^{-2}$ \citep[e.g.,][]{Gall02}.
Although the $N_H$ derived from UV absorption lines cannot account
for the absorption in the X-ray, the UV and X-ray absorbers are
probably closely related \citep*{Bran00}.

Radio-loud quasars are factors of 2--3 times brighter in the X-rays
than radio-quiet quasars with the same optical magnitude
\citep{Brin00}, and tend to have harder X-ray spectra
\citep[$\Gamma\sim 1.6$, e.g.,][]{ReeTur00,Page05}.  These factors
may make them particularly suitable for initial exploratory studies of
the intrinsic X-ray properties and the properties of the line-of-sight 
absorbers \citep{Brot05}.  We report the results of a short 
XMM-Newton observation 
of \balobj\ in this paper.   We detect the object with enough counts to 
compute a hardness ratio, but not enough for more detailed spectral analysis.
Still, the detection can rule out some simple models and has shown us
that even the most powerful radio-loud BAL quasars are weak in X-rays.

\section{X-Ray Observations and Data Analysis}

\begin{figure}
\epsscale{0.4}
\plotone{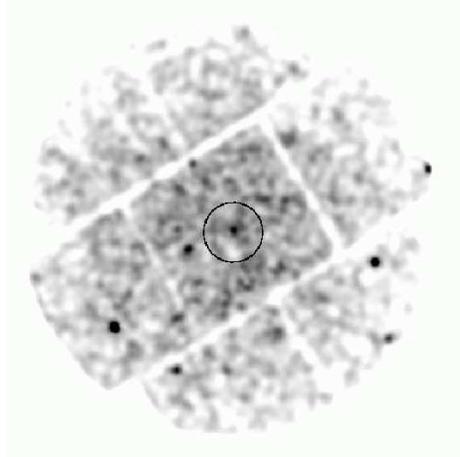}
\caption{FIRST J101614.3+520916 detected by the \xmmn\ MOS2 detector.
\label{fg:xmm}}
\end{figure}

We observed FIRST J1016+5209 with \xmmn\ on November 3, 2001 with a
duration of $\approx$ 10 ks. This was the first X-ray observation of
an FR II BAL quasar.  Unfortunately high X-ray background flares
limited the usable data of the detectors to only 6.1 ks from the EPIC
MOS 1 detector, 5.8 ks from the EPIC MOS 2 detector
(Fig.~\ref{fg:xmm}), and nothing from the more sensitive EPIC-pn
detector.  Table~\ref{tb:xray} gives the X-ray properties of FIRST
J1016+5209.  All of the counts have been background subtracted.  We
define the soft X-ray band to be 0.2--2 keV, and the hard X-ray band
to be 2--8 keV.  The hardness ratio is then determined to be HR =
(H$-$S)/(H+S) = ${-}0.5\pm0.08$, where H and S are the source counts
in the hard and soft bands, respectively, with errors following
\citet{Gehr86}.  Assuming only Galactic absorption, we used PIMMS to
estimate that a photon index of 1.76 would give the
measured hardness ratio seen in the MOS detectors.  There were too few 
counts (46) for a detailed spectral analysis, but the hardness ratio 
indicates an excess of soft photons over hard photons.  

The 0.2--8 keV flux after Galactic absorption correction,
$F_x=6.5\times$10$^{-14}$\,\ergs\ cm$^{-2}$, is calculated using PIMMS
and assuming the average power-law photon index $\Gamma=1.7$ for
radio-loud quasars, consistent with our measured HR and \cite{ReeTur00,Page05}.
We also estimate the rest-frame optical-X-ray spectral index,
$\alpha_{ox}$=$-1.06$, using an optical flux at rest-frame 2500\AA,
and an unabsorbed rest-frame 2\kev\ flux (0.579 keV in the observed frame). 
The unabsorbed rest-frame 2\kev\ flux was calculated using the observed count 
rate, PIMMS and the Galactic absorption, $\Gamma$ = 1.7, and making a 
K-correction.  Cosmological effects in the conversion between fluxes in 
observed frame and rest frame have been taken into account.

While $\alpha_{ox}$=$-1.06$ would indicate a rather X-ray bright
BAL quasar, two additional facts should be considered in 
evaluating the intrinsic X-ray brightness of \balobj: the 
optical flux appears significantly reddened and 
the X-ray brightness may also be estimated based on the radio flux.

We estimate the intrinsic X-ray flux of \balobj\ using the 
radio-X-ray correlation \citep{Brin00} considering
the 3$\sigma$ uncertainty about the correlation\footnote{Due to the 
scatter in the correlation, the uncertainty in
the estimated X-ray flux can be as large as a factor of 1.24.
However, this does not affect $\alpha_{ox}$ very much, because
$\alpha_{ox}$ spans a large frequency range and an uncertainty of a
factor of 2 in X-ray flux only changes $\alpha_{ox}$ by 0.12.}.
We used the total 5 GHz flux from Gregg et al. (2000) and the 
relationship shown in Figure 13 of Brinkmann et al. (2000) for
radio-loud quasars.  The intrinsic X-ray flux is estimated to be 
17 ($\pm21$) times larger than the observed flux in the
ROSAT bandpass.  Based on this apparent supression and an optical flux 
dereddened for intrinsic reddening \citep[see][]{Greg00}, we calculate 
an intrinsic optical-X-ray spectral index, $\alpha_{ox}=-1.19$.
%(cf. the observed $\alpha_{ox}=-1.06$).  
At 2 keV and $z=2.455$, a neutral HI column
density of N$_{H}$ = $8\times10^{23}$ cm$^{-2}$ would be required to
account for the faintness of the observed X-ray flux.  However, such a
high HI column density would result in an extreme hardness ratio, close
to unity, inconsistent with our observed HR=$-0.5$, which is only consistent 
for a column density of N$_{H}$ $\leq$ $1\times$10$^{21.5}$cm$^{-2}$ or less
(assuming a normal radio-loud quasar X-ray slope).  In other words, most of
the observed soft X-ray photons would have been absorbed if X-ray
source in \balobj\ is fully covered by such a high column density absorber.
Therefore, we conclude that the absorber for \balobj\ is 
not a simple neutral absorber with a high column density.
This conclusion should be be tempered by the significant uncertainties in these
estimates, but is consistent with what is seen in other BAL quasars.

\section{Discussion}

As mentioned above, the apparently low X-ray flux and the fact that the
spectrum is not excessively hard together suggest that a fully 
covering neutral absorber with high column density cannot explain our data.  
Possible alternative scenarios for our observed X-rays include a 
partially covering neutral absorber, reflection by an ionized mirror, an 
ionized absorber, or jet contributions.

A partially covering neutral absorber with very high column density
would leave the observed X-ray spectrum similar to the incident
spectrum except for suppressed X-ray flux.  If our estimate of the
intrinsic X-ray flux is correct, the covering factor derived from the
X-ray reduction factor of 17 for \balobj\ would be 94\%.  

An X-ray spectrum dominated by reflection off an ionized ``mirror''
\citep{RosFab93,Ball01} could also explain our data, depending on the
ionization state of the mirror.  In this scenario, at some ionization
parameters, Fe K$\alpha$ emission would present in the X-ray spectrum,
but would require better X-ray observations to be detected.

Ionized absorbers have also often been invoked to explain the X-ray
observations of active galactic nuclei
\citep[e.g.,][]{Kasp02,Grup03,Gall02,Gall04}, since these absorbers can
also be transparent for soft X-ray photons, but our data set has too few
counts to identify any possible absorption edges in order to test 
this explanation.  

Due to the radio-loud nature and the lobe-dominated morphology of this
object, it is also possible that at least part of the observed X-ray
is from the jets.  Recent high-resolution X-ray observations have made
it possible to systematically study X-ray jets and lobes
\citep{Samb02,Samb04,Mars05,Cros05}.  The detection rate is
typically $\sim$60\% \citep{Samb04,Mars05}.  We therefore
speculate that X-rays from the accretion disk could be
completely absorbed, and we are detecting intrinsically weaker but 
unabsorbed X-rays from the
jets, even if beaming effects are not be large given that \balobj\ has a
steep radio spectrum.  The average photon index of the jets is
$\sim$1.5 for a sample of mostly FR~II objects \citep{Samb04}, and the
core-to-jet X-ray flux ratio has a wide range for the detections in
another sample \citep{Mars05}, from 5 to about 200.  This range covers
the supression factor of this object (17) and most of those radio-loud BAL
quasars (42-348) in \citet{Brot05}, which have X-ray fluxes consistent
with what might be expected arising solely in the jets.  Again, 
better data are required to test this explanation.

Finally, there is a possibility that \balobj\ is intrinsically X-ray
faint, or in a low state at the time of the observation since some BAL
quasars do show significant variability \citep[e.g,][]{Gall04}.
Unfortunately our short, high background observation constitutes more of a 
detection rather than a light curve, preventing us from detecting variability.
However, all BAL quasars so far observed with enough counts for
spectral analysis \citep{Gall01} indicate that X-ray absorption is more
likely the primary cause of the ``X-ray weak'' quasars \citep{Laor97}.  

Recently, \citet{Brot05} reported X-ray detections of 5
radio-loud core-dominated BAL quasars with Chandra.  
The hardness ratio ranges from
$-$0.7 to 0.1, and $\alpha_{ox}$ from $-$0.8 to $-$2.0.  All 5 objects
also show significant X-ray suppression compared to estimates of their
intrinsic X-ray flux. Compared with this
sample, \balobj\ does not seem to be abnormal in $\alpha_{ox}$ or 
hardness ratio; our XMM hardness ratio from table 2 (HR =$-0.5$) is equivalent
to a Chandra HR = $-$0.7 (estimated using PIMMS, set to CXO3).   
The X-ray properties of these radio-loud BAL quasars
are in general agreement with the results for radio-quiet BAL
quasars.  Based on their X-ray spectral analyses, \citet{Gall02}
suggested that radio-quiet BAL quasars have typical intrinsic
power-law X-ray continuum of normal radio-quiet quasars, but with
significant absorption column density.  However, they argue that the
absorption is likely very complicated and it is not typically possible to
distinguish between a partially covering and an ionized absorber with
their data.  \citet{Grup03} showed excess soft X-ray photons in their
spectra of 2 radio-quiet BAL quasars, and also reported that both a
partially covering absorber and an ionized absorber could fit their
observed spectra.  It is still not clear whether radio-loud BAL
quasars also have the typical intrinsic X-ray continuum of normal 
radio-loud quasars or whether the FR~II type BAL quasars like \balobj\ 
have special X-ray properties.  High-quality X-ray spectra are needed to 
answer these questions and to reveal the real X-ray nature of \balobj\ 
and other radio-loud BAL quasars.

\section{Conclusions}

We have observed and detected the first confirmed radio-loud FR~II BAL
quasar FIRST J1016+5209 in the X-ray with XMM-Newton for the first time.
We have enough counts to derive the hardness ratio, but not enough for
detailed spectral analysis.  The X-ray flux appears to be suppressed
by a factor of 17 relative to the intrinsic X-rays estimated from the
radio-X-ray correlation, although significant uncertainties are associated
with this factor.  If the X-rays are suppressed due to absorption
associated with a high column density of neutral hydrogen, the
X-rays observed would be much harder, which is inconsistent with the
observations.  This implies that the X-ray absorption in \balobj\ is
more complicated, such as an ionized absorber, an ionized mirror, or
neutral but partially covering the X-ray source.  Contributions from
a jet are also possible.  High-quality X-ray spectra are necessary to 
understand the nature of the absorber.

\acknowledgments

This work is funded by Wyoming NASA Space Grant Consortium, NASA Grant
NGT-40102 and by Wyoming NASA EPSCoR NASA Grant NCC5-578.
This work is also funded in part by NASA through the US XMM-Newton
Program with data provided by ESA.  This work was partly performed
under the auspices of the US Department of
Energy by the University of California, Lawrence Livermore National
Laboratory under contract No. W-7405-Eng-48.

%\clearpage

%\clearpage

%%%%%%%%%%%%%%%%%%%%%%
\begin{deluxetable}{ccccccccc}
\tabletypesize{\scriptsize}
\tablecolumns{9}
\tablewidth{0pc}
\tablecaption{Optical and Ratio Properties of \balobj\ \label{tb:para}}
\tablehead{
{BAL Quasar}  & $z$  & E ($\approx$R)  & 
S$_{{\rm 20 cm}}$  & $A_V$\tablenotemark{a}  & M$_{B}$  & 
log($L_{\rm 5 GHz}$)  & log($R^\ast$) &
f$_{2500\mbox{\AA}}$\tablenotemark{a}\\ 
        & & (Mag)  & mJy  & (Mag)  & (Mag)  & 
(ergs s$^{-1}$Hz$^{-1}$)  & & 
(10$^{-29}$ergs s$^{-1}$cm$^{-2}$Hz$^{-1}$)
}
\startdata
J1016+5209  & 2.455  & 18.6  & 177  & 0.35  & 
$-26.2\,(-27.3)$  & 34.3  & 3.4 (2.7)  &8.24 (17.3)
\enddata
\tablecomments{
Parameters from \citet{Greg00} unless noted.  
$A_V$ indicates the intrinsic reddening 
estimated by Small Magellanic Cloud reddening law \citep{Prev84}
and matching the UV spectrum of \balobj\ to the FBQS composite quasar
spectrum \citep{Brot01}.  Galactic reddening in this direction
is insignificant ($A_V=0.017$).  $L_{5 GHz}$ and $R^\ast$ (ratio of
radio-optical brightness) are for the total radio flux, including that 
of both the core and lobes.  Values in the parentheses
have been corrected for intrinsic reddening.  We note that the absolute 
magnitude was k-corrected by \citet{Greg00} based on the broad-band colors.
}
\tablenotetext{a}{Not from \citet{Greg00}.  Calculated for this work.
The values of f$_{2500\mbox{\AA}}$ are in observed frame.
}
\end{deluxetable}

\begin{table}
\caption{X-Ray Properties of \balobj\ \label{tb:xray}}
\scriptsize
\begin{tabular}{ccccccccc} \hline \hline
{BAL Quasar}  & $N_{H}$  & Counts s$^{-1}$  & 
Soft  & Hard  & S+H  & HR  & F$_{X}$  & 
$\alpha_{OX}$ \\
& (cm$^{-2}$)   & ($10^{-4}$)   & cts   & 
    cts    & cts   & & 
(ergs s$^{-1}$cm$^{-2}$)   & \\
(1) & (2) & (3) & (4) & (5) & (6) & (7) & (8) & (9)
\\ \hline 
J1016+5209  & $7.64\times 10^{19}$  & $77\pm6$  & $36\pm2.4$  & 
    $10\pm1.3$  & $46\pm3.6$  & $-0.5\pm0.08$  &
$(6.5\pm0.7)\times10^{-14}$  & 
    $-1.06\,(-1.19)$ 
\\ \hline
\end{tabular}
\normalsize
\tablecomments{
Col. (2): The Galactic neutral hydrogen column density
\citep{DicLoc90}.
Col. (3): X-ray counts per second (0.2--8 \kev) from the two MOS detectors.
Col. (4): Counts in the soft bandpass (S, 0.2--2 \kev) from the two MOS detectors.
Col. (5): Counts in the hard bandpass (H, 2--8 \kev) from the two MOS detectors.
Col. (6): Total counts (0.2--8 \kev) from the two MOS detectors.
Col. (7): Hardness ratio defined as (H-S)/(H+S), error following
\citet{Gehr86}.
Col. (8): The observed, unabsorbed 0.2--8 \kev\ X-ray flux using PIMMS and
assuming only the Galactic column density and the photon index
$\Gamma=1.7$.
Col. (9): The optical-X-ray spectral index (rest frame 2500\AA--2\kev).  The value in the
parentheses is calculated from estimated intrinsic X-ray flux and
a dereddened optical flux.
}
\end{table}

\end{document}